\documentclass[aps,pre,twocolumn,superscriptaddress,longbibliography,hidelinks]{revtex4-2}

\usepackage{color}
\usepackage{graphicx}
\usepackage{siunitx}
\usepackage{amsmath}
\usepackage{hyperref}

\begin{document}

\title{Oscillatory collective motion in viscoelastic and elastic active fluids and solids\\ under circular confinement}

\author{Henning Reinken}
\email{henning.reinken@ovgu.de}
\affiliation{Institut f\"ur Physik, Otto-von-Guericke-Universität Magdeburg, Universitätsplatz 2, 39106 Magdeburg, Germany} 

\author{Andreas M. Menzel}
\email{a.menzel@ovgu.de}
\affiliation{Institut f\"ur Physik, Otto-von-Guericke-Universität Magdeburg, Universitätsplatz 2, 39106 Magdeburg, Germany} 

\begin{abstract}
In an inspiring recent study, Xu \textit{et al.} [Nat. Phys. \textbf{19}, 46 (2023)] observed for a living active biofilm under circular confinement two emergent dynamic modes of collective motion in the film. One corresponds to global rotational motion of oscillating sense of rotation, the other one to uniformly translating motion of rotating migration direction. The authors reproduced these features in a discretized theoretical model for elastic active solids. We here demonstrate that the discovered fundamental phenomena are generic and emerge abundantly for a broad range of viscoelastic fluids and solids. Elastic solids represent only one limiting case. 
\end{abstract}

\date{\today}

\maketitle

Recently, Xu \textit{et al.}~\cite{xu2023autonomous} performed a pioneering experiment on \textit{Proteus mirabilis} biofilms. It represents one of the first experimental realizations of living solids for the purpose of studying different types of persistent motion on large collective scales. %
Corresponding properties of active solids were previously observed mainly from experiments on small artificial active robots~\cite{ferrante2012self,shen2016probing,baconnier2022selective,zheng2022experimental}. 

In the considered biofilms, the constituent bacteria are embedded in an extracellular polymeric matrix. The latter gives rise to elastic interactions between the individuals.
Specifically, under circular confinement Xu \textit{et al.} observe the emergence of two distinct oscillatory global modes of motion. They are selectively excited depending on the activity of the bacteria~\cite{xu2023autonomous}. 
While at low activity, the dynamics can be described
as an oscillatory rotational motion, at high activity, a translational mode of rotating velocity direction emerges. 

Because of the elastic contributions to the bacterial interactions resulting from the extracellular matrix, the films were conceived as elastic solids. Accordingly, theoretical analysis was based on lattice-like configurations of elastically coupled active agents linked by permanently set elastic springs~\cite{ferrante2013elasticity,scheibner2020odd}. 
Since the results are reproduced by the elastic spring-lattice model, the observed properties appear associated with genuinely solid-like behavior. 

However, it turns out that the behavior discovered by Xu \textit{et al.} is much more generic. We here show that ultimate solid-like behavior is not mandatory to observe the described modes of motion. Instead, viscoelastic fluids that feature terminal flow instead of elastic arrest likewise display the corresponding behavior. Therefore, the modes of motion discovered by Xu \textit{et al.} should be characteristic for a broad general class of biological fluids, solids, and biomaterials. 
Active elastic solids only represent one limiting case.

To illustrate these conclusions, we apply a recently derived, unified description of migrating active units in viscous, viscoelastic, or elastic active media on a supporting substrate~\cite{reinken2025unified}.
The behavior of the active medium is represented by the dynamics of three continuous fields.
First, the polar order parameter field $\mathbf{P}$ quantifies the locally averaged polar orientational order of the migration directions of the active units.
Second, the overall velocity field $\mathbf{v}$ couples to the dynamics of $\mathbf{P}$.
Third, to take elastic properties into account, we capture the deformation history of the system by the displacement field $\mathbf{u}$.

As shown in Ref.~\onlinecite{reinken2025unified}, the basic evolution of $\mathbf{P}$ in the simplest case in rescaled units is governed by
\begin{eqnarray}
\partial_{{t}} \mathbf{P} + {\mathbf{v}} \cdot {\nabla} \mathbf{P} &= & {}- \mathbf{P} + {\nabla}^2 \mathbf{P} + {\boldsymbol{\Omega}} \cdot \mathbf{P}
\nonumber\\ 
&& {}+ {\gamma}_\mathrm{a} {\mathbf{v}} \cdot [(2 + \mathbf{P}\cdot\mathbf{P})\mathbf{I}/3 - \mathbf{P}\mathbf{P}] \, .
\label{eq:EvolutionEquationPolarOrderRescaled}
\end{eqnarray}
The first two terms on the right-hand side reflect orientational and translational diffusion.
This equation further includes advection with the flow field $\mathbf{v}$ as well as reorientation via the vorticity tensor $\boldsymbol{\Omega} = [(\nabla \mathbf{v})^\top - (\nabla \mathbf{v})]$, where $^\top$ denotes the transpose.
Moreover, the active units tend to align with the overall flow, quantified by the alignment strength $\gamma_\mathrm{a}$.
This %
effect is specific to active units interacting front--rear asymmetrically with a supporting substrate and exposed to displacements of a surrounding medium~\cite{brotto2013hydrodynamics,maitra2020swimmer}. 
Similar terms are included in various particle-resolved models of active solids~\cite{baconnier2022selective,baconnier2024self}.
Further, $\mathbf{I}$ is the unit matrix, and the magnitude of the vector field of polar order is limited to %
$|\mathbf{P}|\leq 1$.

The velocity field $\mathbf{v}$ is governed by an extended Stokes equation,
\begin{equation}
\label{eq:StokesEquationRescaled}
\mathbf{0} = - {\nabla} {p} + {\eta} {\nabla}^2 {\mathbf{v}}  + {\mu}{\nabla}^2 {\mathbf{u}} - {\nu}_\mathrm{v} {\mathbf{v}} - {\nu}_\mathrm{d} {\mathbf{u}} + {\nu}_\mathrm{p} \mathbf{P} \, .
\end{equation}
Here, the pressure field $p$ enforces incompressibility, that is, $\nabla \cdot \mathbf{v} = 0$. %
Viscous and elastic force densities are quantified by the viscosity $\eta$ and shear modulus $\mu$, respectively, together with the displacement field $\mathbf{u}$.
Motion relative to the substrate implies friction and elastic restoring forces quantified by $\nu_\mathrm{v}$ and $\nu_\mathrm{d}$, respectively. They slow down motion and tend to pull material elements back to their initial positions.
Active forcing of strength $\nu_\mathrm{p}$ %
drive the film. 

\begin{figure*}
\includegraphics[width=0.999\linewidth]{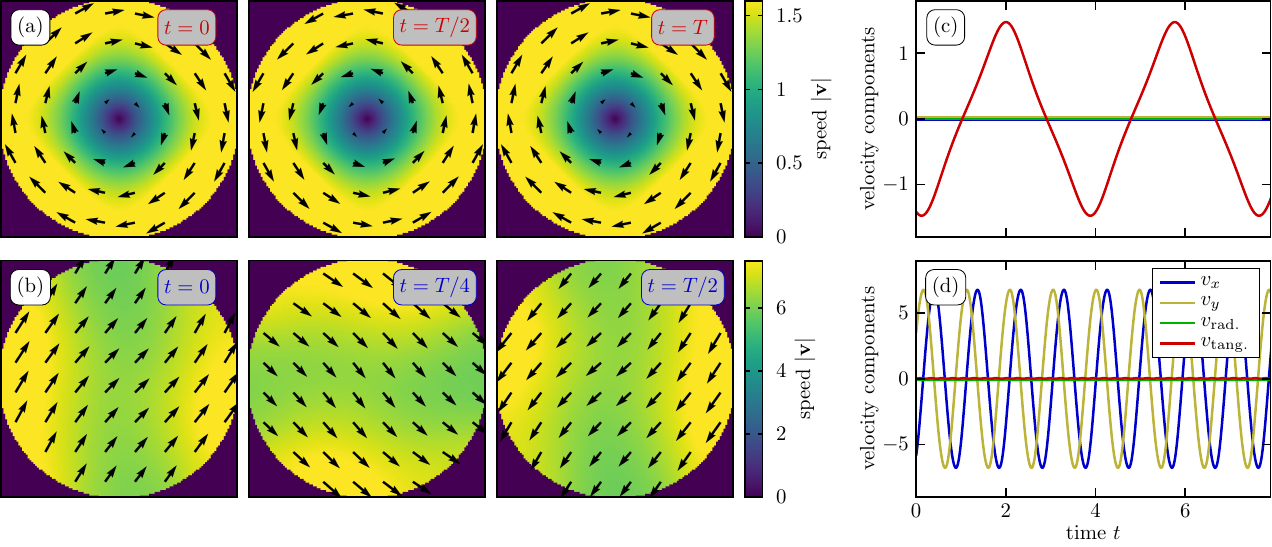}
\caption{\label{fig:ModesOscConf}Emergent oscillatory modes of active collective motion under circular confinement for a viscoelastic fluid. (a) and (b) show snapshots of the overall velocity field at different instances during the oscillation period $T$. (a) At lower activity, $\nu_\mathrm{p} = 22$, the oscillatory rotational mode emerges, while (b) at higher activity, $\nu_\mathrm{p} = 23$, the translational mode of globally rotating uniform velocity is observed. 
Arrows and color indicate directions and magnitude of the velocity field, respectively. (c) and (d) display the time evolution of the spatially averaged $x$-, $y$-, radial, and tangential components of the velocity field for (c) the oscillatory rotational mode and (d) the rotating translational mode. The two modes are clearly distinguished from each other by complementary vanishing of the different velocity components. This active viscoelastic example fluid is characterized by parameter values $\tau_\mathrm{d} = 1$, $\gamma_\mathrm{a} = 1$, $\eta = 1$, $\mu = 1$, $\nu_\mathrm{v} = 1$, and $\nu_\mathrm{d} = 10$.
}
\end{figure*}

\begin{figure}
\includegraphics[width=0.999\linewidth]{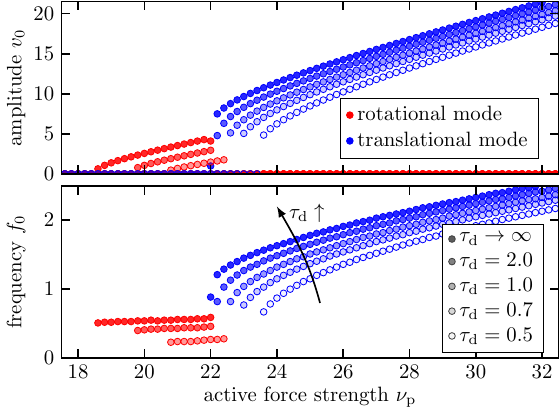}
\caption{\label{fig:FreqOscConf}Transition between oscillatory rotational modes of collective motion and dynamic states of rotating uniform translational migration. (a) Amplitude $v_0$ and (b) frequency $f_0$ of the oscillations of the migration directions as a function of the active driving strength $\nu_\mathrm{p}$ for different relaxation times $\tau_\mathrm{d}$ of the elastic memory. 
The transition from oscillatory rotational to rotating translational motion is clearly observed for a wide range of relaxation times $\tau_\mathrm{d}$ characterizing viscoelastic fluids (finite $\tau_\mathrm{d}$) up to the limit of purely elastic solids ($\tau_\mathrm{d} \to \infty$). The remaining material parameters are set to $\gamma_\mathrm{a} = 1$, $\eta = 1$, $\mu = 1$, $\nu_\mathrm{v} = 1$, and $\nu_\mathrm{d} = 10$.
}
\end{figure}

The dynamics of the displacement field $\mathbf{u}$ is governed by~\cite{puljiz2019memory}
\begin{equation}
\label{eq:DisplacementFieldEvolution}
\partial_{{t}} {\mathbf{u}}  = {\mathbf{v}} - \tau_\mathrm{d}^{-1} \mathbf{u} \, .
\end{equation}
Thus, overall motion $\mathbf{v}$ drives distortions, while the memory of the initial state as quantified by $\mathbf{u}$ decays with a relaxation time $\tau_\mathrm{d}$. The latter process implies net flow and therefore incomplete elasticity. Only the limit $\tau_\mathrm{d} \to \infty$ corresponds to full elastic reversibility in terms of solid-like behavior.
A finite $\tau_\mathrm{d}$ describes a viscoelastic fluid, while $\tau_\mathrm{d}\to0$ refers to a viscous fluid.
Thus, by tuning $\tau_\mathrm{d}$ we cover a whole range of viscoelastic behavior.

To compare the dynamics emerging from this theoretical description of viscoelastic fluids with the experimental observations in Ref.~\onlinecite{xu2023autonomous}, we numerically solve the above equations %
for a flat film under circular confinement.
For this purpose, an elastic restoring force acting normal to the boundary is introduced, in analogy to the investigation by %
Xu \textit{et al.}~\cite{xu2023autonomous}
(see the Supplemental Information for numerical details).

As a central experimental observation, two oscillatory modes emerge, which are controlled by activity~\cite{xu2023autonomous}.
At low activity, the dynamics can be described as an oscillatory rotational motion.
Here, the velocity oscillates back and forth in the circumferential direction. The global mean velocity vanishes. 
At high activity, an approximately uniform translational mode emerges.
While the velocity magnitude remains roughly constant, the direction of motion persistently rotates.
Both effects were observed for active elastic solids in Ref.~\onlinecite{xu2023autonomous}. 

We here highlight that the effects are more generic. Likewise, we observe both modes of global motion for a broad range of viscoelastic fluids, that is, relatively short relaxation times $\tau_\mathrm{d}$ in Eq.~(\ref{eq:DisplacementFieldEvolution}).
Figure~\ref{fig:ModesOscConf} demonstrates this observation, where we show snapshots of the velocity field for the two distinct modes at different instances during the oscillation periods.
Dynamic illustrations are found in the Supplementary Movies.
We set $\tau_\mathrm{d} = 1$, corresponding to a genuinely viscoelastic fluid. In our scaled units, the loss time of the elastic memory %
then is comparable to the reorientation time of active motion.

Figure~\ref{fig:FreqOscConf} shows amplitudes and frequencies of the two modes as a function of the active driving strength $\nu_\mathrm{P}$ for decreasing relaxation times $\tau_\mathrm{d}$ of the elastic memory and thus increasing fluid-like character.
The transition from oscillatory rotational to rotating translational motion at a certain activity is observed for a wide range of values of $\tau_\mathrm{d}$. It starts from  purely elastic solids ($\tau_\mathrm{d} \to \infty$) and extends far into the regime of viscoelastic fluids (finite $\tau_\mathrm{d}$). 
Also for viscoelastic fluids we find  that rotational frequency increases with activity~\cite{xu2023autonomous}.
Here, we further observe that the oscillations become slower for decreasing $\tau_\mathrm{d}$, that is, with increasing fluid-like behavior.
At the same time, the oscillation amplitude decreases.
Moving closer to the limit of viscous fluids, the oscillatory rotational mode vanishes at a certain point (here at $\tau_\mathrm{d} \approx 0.7$).
In these cases, activity outperforms remaining elastic forces and the system develops states of global polar orientational order.
However, translational oscillations at larger activities may still be observed. In Fig.~\ref{fig:FreqOscConf}, this leads to the (blue) curves for the translational mode that do not show a (red) counterpart for the rotational mode for some values of $\tau_\mathrm{d}$.

In conclusion, we demonstrate that the two distinct oscillatory modes of collective motion observed in the inspiring study by Xu \textit{et al.}~\cite{xu2023autonomous} are generic. The observed phenomena extend beyond the regime of active solids and likewise emerge in viscoelastic active fluids.
Concerning our understanding of biofilms, these results promote the picture of biofilms as complex viscoelastic media with remarkable rheological properties.\\

\noindent \textbf{Acknowledgments}\\
\noindent The authors thank the Deutsche Forschungsgemeinschaft (German Research Foundation, DFG) for support through the Research Grant no.\ ME 3571/5-1. Moreover, A.M.M. acknowledges support by the DFG through the Heisenberg Grant no.\ ME 3571/4-1.  \\

\noindent \textbf{Author contributions}\\
\noindent A.M.M.\ and H.R.\ designed the study and wrote the manuscript. H.R.\ developed and performed the computational calculations. \\

\noindent \textbf{Competing interests}\\
\noindent The authors declare no competing interests.

\onecolumngrid

\bibliography{references}

\begin{thebibliography}{12}%
\makeatletter
\providecommand \@ifxundefined [1]{%
 \@ifx{#1\undefined}
}%
\providecommand \@ifnum [1]{%
 \ifnum #1\expandafter \@firstoftwo
 \else \expandafter \@secondoftwo
 \fi
}%
\providecommand \@ifx [1]{%
 \ifx #1\expandafter \@firstoftwo
 \else \expandafter \@secondoftwo
 \fi
}%
\providecommand \natexlab [1]{#1}%
\providecommand \enquote  [1]{``#1''}%
\providecommand \bibnamefont  [1]{#1}%
\providecommand \bibfnamefont [1]{#1}%
\providecommand \citenamefont [1]{#1}%
\providecommand \href@noop [0]{\@secondoftwo}%
\providecommand \href [0]{\begingroup \@sanitize@url \@href}%
\providecommand \@href[1]{\@@startlink{#1}\@@href}%
\providecommand \@@href[1]{\endgroup#1\@@endlink}%
\providecommand \@sanitize@url [0]{\catcode `\\12\catcode `\$12\catcode
  `\&12\catcode `\#12\catcode `\^12\catcode `\_12\catcode `\%12\relax}%
\providecommand \@@startlink[1]{}%
\providecommand \@@endlink[0]{}%
\providecommand \url  [0]{\begingroup\@sanitize@url \@url }%
\providecommand \@url [1]{\endgroup\@href {#1}{\urlprefix }}%
\providecommand \urlprefix  [0]{URL }%
\providecommand \Eprint [0]{\href }%
\providecommand \doibase [0]{https://doi.org/}%
\providecommand \selectlanguage [0]{\@gobble}%
\providecommand \bibinfo  [0]{\@secondoftwo}%
\providecommand \bibfield  [0]{\@secondoftwo}%
\providecommand \translation [1]{[#1]}%
\providecommand \BibitemOpen [0]{}%
\providecommand \bibitemStop [0]{}%
\providecommand \bibitemNoStop [0]{.\EOS\space}%
\providecommand \EOS [0]{\spacefactor3000\relax}%
\providecommand \BibitemShut  [1]{\csname bibitem#1\endcsname}%
\let\auto@bib@innerbib\@empty
\bibitem [{\citenamefont {Xu}\ \emph {et~al.}(2023)\citenamefont {Xu},
  \citenamefont {Huang}, \citenamefont {Zhang},\ and\ \citenamefont
  {Wu}}]{xu2023autonomous}%
  \BibitemOpen
  \bibfield  {author} {\bibinfo {author} {\bibfnamefont {H.}~\bibnamefont
  {Xu}}, \bibinfo {author} {\bibfnamefont {Y.}~\bibnamefont {Huang}}, \bibinfo
  {author} {\bibfnamefont {R.}~\bibnamefont {Zhang}},\ and\ \bibinfo {author}
  {\bibfnamefont {Y.}~\bibnamefont {Wu}},\ }\bibfield  {title} {\bibinfo
  {title} {Autonomous waves and global motion modes in living active solids},\
  }\href@noop {} {\bibfield  {journal} {\bibinfo  {journal} {Nature Phys.}\
  }\textbf {\bibinfo {volume} {19}},\ \bibinfo {pages} {46} (\bibinfo {year}
  {2023})}\BibitemShut {NoStop}%
\bibitem [{\citenamefont {Ferrante}\ \emph {et~al.}(2012)\citenamefont
  {Ferrante}, \citenamefont {Turgut}, \citenamefont {Huepe}, \citenamefont
  {Stranieri}, \citenamefont {Pinciroli},\ and\ \citenamefont
  {Dorigo}}]{ferrante2012self}%
  \BibitemOpen
  \bibfield  {author} {\bibinfo {author} {\bibfnamefont {E.}~\bibnamefont
  {Ferrante}}, \bibinfo {author} {\bibfnamefont {A.~E.}\ \bibnamefont
  {Turgut}}, \bibinfo {author} {\bibfnamefont {C.}~\bibnamefont {Huepe}},
  \bibinfo {author} {\bibfnamefont {A.}~\bibnamefont {Stranieri}}, \bibinfo
  {author} {\bibfnamefont {C.}~\bibnamefont {Pinciroli}},\ and\ \bibinfo
  {author} {\bibfnamefont {M.}~\bibnamefont {Dorigo}},\ }\bibfield  {title}
  {\bibinfo {title} {Self-organized flocking with a mobile robot swarm: {A}
  novel motion control method},\ }\href@noop {} {\bibfield  {journal} {\bibinfo
   {journal} {Adapt. Behav.}\ }\textbf {\bibinfo {volume} {20}},\ \bibinfo
  {pages} {460} (\bibinfo {year} {2012})}\BibitemShut {NoStop}%
\bibitem [{\citenamefont {Shen}\ \emph {et~al.}(2016)\citenamefont {Shen},
  \citenamefont {Tan},\ and\ \citenamefont {Xu}}]{shen2016probing}%
  \BibitemOpen
  \bibfield  {author} {\bibinfo {author} {\bibfnamefont {H.}~\bibnamefont
  {Shen}}, \bibinfo {author} {\bibfnamefont {P.}~\bibnamefont {Tan}},\ and\
  \bibinfo {author} {\bibfnamefont {L.}~\bibnamefont {Xu}},\ }\bibfield
  {title} {\bibinfo {title} {Probing the role of mobility in the collective
  motion of nonequilibrium systems},\ }\href@noop {} {\bibfield  {journal}
  {\bibinfo  {journal} {Phys. Rev. Lett.}\ }\textbf {\bibinfo {volume} {116}},\
  \bibinfo {pages} {048302} (\bibinfo {year} {2016})}\BibitemShut {NoStop}%
\bibitem [{\citenamefont {Baconnier}\ \emph {et~al.}(2022)\citenamefont
  {Baconnier}, \citenamefont {Shohat}, \citenamefont {L{\'o}pez}, \citenamefont
  {Coulais}, \citenamefont {D{\'e}mery}, \citenamefont {D{\"u}ring},\ and\
  \citenamefont {Dauchot}}]{baconnier2022selective}%
  \BibitemOpen
  \bibfield  {author} {\bibinfo {author} {\bibfnamefont {P.}~\bibnamefont
  {Baconnier}}, \bibinfo {author} {\bibfnamefont {D.}~\bibnamefont {Shohat}},
  \bibinfo {author} {\bibfnamefont {C.~H.}\ \bibnamefont {L{\'o}pez}}, \bibinfo
  {author} {\bibfnamefont {C.}~\bibnamefont {Coulais}}, \bibinfo {author}
  {\bibfnamefont {V.}~\bibnamefont {D{\'e}mery}}, \bibinfo {author}
  {\bibfnamefont {G.}~\bibnamefont {D{\"u}ring}},\ and\ \bibinfo {author}
  {\bibfnamefont {O.}~\bibnamefont {Dauchot}},\ }\bibfield  {title} {\bibinfo
  {title} {Selective and collective actuation in active solids},\ }\href@noop
  {} {\bibfield  {journal} {\bibinfo  {journal} {Nature Phys.}\ }\textbf
  {\bibinfo {volume} {18}},\ \bibinfo {pages} {1234} (\bibinfo {year}
  {2022})}\BibitemShut {NoStop}%
\bibitem [{\citenamefont {Zheng}\ \emph {et~al.}(2022)\citenamefont {Zheng},
  \citenamefont {Huepe},\ and\ \citenamefont {Han}}]{zheng2022experimental}%
  \BibitemOpen
  \bibfield  {author} {\bibinfo {author} {\bibfnamefont {Y.}~\bibnamefont
  {Zheng}}, \bibinfo {author} {\bibfnamefont {C.}~\bibnamefont {Huepe}},\ and\
  \bibinfo {author} {\bibfnamefont {Z.}~\bibnamefont {Han}},\ }\bibfield
  {title} {\bibinfo {title} {Experimental capabilities and limitations of a
  position-based control algorithm for swarm robotics},\ }\href@noop {}
  {\bibfield  {journal} {\bibinfo  {journal} {Adapt. Behav.}\ }\textbf
  {\bibinfo {volume} {30}},\ \bibinfo {pages} {19} (\bibinfo {year}
  {2022})}\BibitemShut {NoStop}%
\bibitem [{\citenamefont {Ferrante}\ \emph {et~al.}(2013)\citenamefont
  {Ferrante}, \citenamefont {Turgut}, \citenamefont {Dorigo},\ and\
  \citenamefont {Huepe}}]{ferrante2013elasticity}%
  \BibitemOpen
  \bibfield  {author} {\bibinfo {author} {\bibfnamefont {E.}~\bibnamefont
  {Ferrante}}, \bibinfo {author} {\bibfnamefont {A.~E.}\ \bibnamefont
  {Turgut}}, \bibinfo {author} {\bibfnamefont {M.}~\bibnamefont {Dorigo}},\
  and\ \bibinfo {author} {\bibfnamefont {C.}~\bibnamefont {Huepe}},\ }\bibfield
   {title} {\bibinfo {title} {Elasticity-based mechanism for the collective
  motion of self-propelled particles with springlike interactions: {A} model
  system for natural and artificial swarms},\ }\href@noop {} {\bibfield
  {journal} {\bibinfo  {journal} {Phys. Rev. Lett.}\ }\textbf {\bibinfo
  {volume} {111}},\ \bibinfo {pages} {268302} (\bibinfo {year}
  {2013})}\BibitemShut {NoStop}%
\bibitem [{\citenamefont {Scheibner}\ \emph {et~al.}(2020)\citenamefont
  {Scheibner}, \citenamefont {Souslov}, \citenamefont {Banerjee}, \citenamefont
  {Sur{\'o}wka}, \citenamefont {Irvine},\ and\ \citenamefont
  {Vitelli}}]{scheibner2020odd}%
  \BibitemOpen
  \bibfield  {author} {\bibinfo {author} {\bibfnamefont {C.}~\bibnamefont
  {Scheibner}}, \bibinfo {author} {\bibfnamefont {A.}~\bibnamefont {Souslov}},
  \bibinfo {author} {\bibfnamefont {D.}~\bibnamefont {Banerjee}}, \bibinfo
  {author} {\bibfnamefont {P.}~\bibnamefont {Sur{\'o}wka}}, \bibinfo {author}
  {\bibfnamefont {W.~T.}\ \bibnamefont {Irvine}},\ and\ \bibinfo {author}
  {\bibfnamefont {V.}~\bibnamefont {Vitelli}},\ }\bibfield  {title} {\bibinfo
  {title} {Odd elasticity},\ }\href@noop {} {\bibfield  {journal} {\bibinfo
  {journal} {Nature Phys.}\ }\textbf {\bibinfo {volume} {16}},\ \bibinfo
  {pages} {475} (\bibinfo {year} {2020})}\BibitemShut {NoStop}%
\bibitem [{\citenamefont {Reinken}\ and\ \citenamefont
  {Menzel}(2025)}]{reinken2025unified}%
  \BibitemOpen
  \bibfield  {author} {\bibinfo {author} {\bibfnamefont {H.}~\bibnamefont
  {Reinken}}\ and\ \bibinfo {author} {\bibfnamefont {A.~M.}\ \bibnamefont
  {Menzel}},\ }\bibfield  {title} {\bibinfo {title} {Unified description of
  viscous, viscoelastic, or elastic thin active films on substrates},\
  }\href@noop {} {\bibfield  {journal} {\bibinfo  {journal} {arXiv preprint
  arXiv:2502.04802}\ } (\bibinfo {year} {2025})}\BibitemShut {NoStop}%
\bibitem [{\citenamefont {Brotto}\ \emph {et~al.}(2013)\citenamefont {Brotto},
  \citenamefont {Caussin}, \citenamefont {Lauga},\ and\ \citenamefont
  {Bartolo}}]{brotto2013hydrodynamics}%
  \BibitemOpen
  \bibfield  {author} {\bibinfo {author} {\bibfnamefont {T.}~\bibnamefont
  {Brotto}}, \bibinfo {author} {\bibfnamefont {J.-B.}\ \bibnamefont {Caussin}},
  \bibinfo {author} {\bibfnamefont {E.}~\bibnamefont {Lauga}},\ and\ \bibinfo
  {author} {\bibfnamefont {D.}~\bibnamefont {Bartolo}},\ }\bibfield  {title}
  {\bibinfo {title} {Hydrodynamics of confined active fluids},\ }\href@noop {}
  {\bibfield  {journal} {\bibinfo  {journal} {Phys. Rev. Lett.}\ }\textbf
  {\bibinfo {volume} {110}},\ \bibinfo {pages} {038101} (\bibinfo {year}
  {2013})}\BibitemShut {NoStop}%
\bibitem [{\citenamefont {Maitra}\ \emph {et~al.}(2020)\citenamefont {Maitra},
  \citenamefont {Srivastava}, \citenamefont {Marchetti}, \citenamefont
  {Ramaswamy},\ and\ \citenamefont {Lenz}}]{maitra2020swimmer}%
  \BibitemOpen
  \bibfield  {author} {\bibinfo {author} {\bibfnamefont {A.}~\bibnamefont
  {Maitra}}, \bibinfo {author} {\bibfnamefont {P.}~\bibnamefont {Srivastava}},
  \bibinfo {author} {\bibfnamefont {M.~C.}\ \bibnamefont {Marchetti}}, \bibinfo
  {author} {\bibfnamefont {S.}~\bibnamefont {Ramaswamy}},\ and\ \bibinfo
  {author} {\bibfnamefont {M.}~\bibnamefont {Lenz}},\ }\bibfield  {title}
  {\bibinfo {title} {Swimmer suspensions on substrates: {A}nomalous stability
  and long-range order},\ }\href@noop {} {\bibfield  {journal} {\bibinfo
  {journal} {Phys. Rev. Lett.}\ }\textbf {\bibinfo {volume} {124}},\ \bibinfo
  {pages} {028002} (\bibinfo {year} {2020})}\BibitemShut {NoStop}%
\bibitem [{\citenamefont {Baconnier}\ \emph {et~al.}(2024)\citenamefont
  {Baconnier}, \citenamefont {Dauchot}, \citenamefont {D{\'e}mery},
  \citenamefont {D{\"u}ring}, \citenamefont {Henkes}, \citenamefont {Huepe},\
  and\ \citenamefont {Shee}}]{baconnier2024self}%
  \BibitemOpen
  \bibfield  {author} {\bibinfo {author} {\bibfnamefont {P.}~\bibnamefont
  {Baconnier}}, \bibinfo {author} {\bibfnamefont {O.}~\bibnamefont {Dauchot}},
  \bibinfo {author} {\bibfnamefont {V.}~\bibnamefont {D{\'e}mery}}, \bibinfo
  {author} {\bibfnamefont {G.}~\bibnamefont {D{\"u}ring}}, \bibinfo {author}
  {\bibfnamefont {S.}~\bibnamefont {Henkes}}, \bibinfo {author} {\bibfnamefont
  {C.}~\bibnamefont {Huepe}},\ and\ \bibinfo {author} {\bibfnamefont
  {A.}~\bibnamefont {Shee}},\ }\bibfield  {title} {\bibinfo {title}
  {Self-aligning polar active matter},\ }\href@noop {} {\bibfield  {journal}
  {\bibinfo  {journal} {arXiv preprint arXiv:2403.10151}\ } (\bibinfo {year}
  {2024})}\BibitemShut {NoStop}%
\bibitem [{\citenamefont {Puljiz}\ and\ \citenamefont
  {Menzel}(2019)}]{puljiz2019memory}%
  \BibitemOpen
  \bibfield  {author} {\bibinfo {author} {\bibfnamefont {M.}~\bibnamefont
  {Puljiz}}\ and\ \bibinfo {author} {\bibfnamefont {A.~M.}\ \bibnamefont
  {Menzel}},\ }\bibfield  {title} {\bibinfo {title} {Memory-based mediated
  interactions between rigid particulate inclusions in viscoelastic
  environments},\ }\href@noop {} {\bibfield  {journal} {\bibinfo  {journal}
  {Phys. Rev. E}\ }\textbf {\bibinfo {volume} {99}},\ \bibinfo {pages} {012601}
  (\bibinfo {year} {2019})}\BibitemShut {NoStop}%
\end{thebibliography}%


%

\end{document}


\title{Oscillatory collective motion in viscoelastic and elastic active fluids and solids \\under
circular confinement\\[1\baselineskip] \textit{Supplementary Information}}

\author{Henning Reinken}
\email{henning.reinken@ovgu.de}
\affiliation{Institut f\"ur Physik, Otto-von-Guericke-Universität Magdeburg, Universitätsplatz 2, 39106 Magdeburg, Germany} 

\author{Andreas M. Menzel}
\email{a.menzel@ovgu.de}
\affiliation{Institut f\"ur Physik, Otto-von-Guericke-Universität Magdeburg, Universitätsplatz 2, 39106 Magdeburg, Germany} 

\begin{abstract}
    This Supplementary Information overviews the technical background of our numerical simulations. We perform them to solve the listed equations of motion in the main article to derive the presented results. 
\end{abstract}

\date{\today}

\maketitle

To compare the results derived from our dynamic equations of motion with the experimental observations by Xu \textit{et al.}~\cite{xu2023autonomous}, we numerically solve Eqs.~(1), (2), and (3) listed in the main text for a flat two-dimensional system under circular confinement.
We employ a finite-difference scheme to numerically resolve the spatial derivatives.
For discretization of the Laplacian, we use the isotropic Oono-Puri stencil, which incorporates 9 grid points~\cite{provatas2011phase}.
To significantly speed up the calculations, we rely on a sparse-matrix implementation within the framework of SciPy~\cite{2020SciPy}.
We employ Neumann boundary conditions, that is, the gradients of all quantities vanish in the normal direction at the circular boundary of the system.
Similarly to Ref.~\cite{xu2023autonomous}, we add an additional elastic restoring force that only acts on the  component of the displacement field that is normal to the boundary of the circular domain to describe the steric effects of the circular confinement.

The equations for low-Reynolds-number dynamics of incompressible fluids, see Eq.~(2) in the main text, thus are supplemented as
\begin{equation}
\label{eq:StokesEquationPlusConfinement}
\begin{aligned}
0&= \nabla \cdot \mathbf{v}\, , \\ 
\mathbf{0}&= - \nabla p + \eta \nabla^2 \mathbf{v} + \mu \nabla^2 \mathbf{u} - \nu_\mathrm{v}  \mathbf{v} - \nu_\mathrm{d}  \mathbf{u} - \nu_\mathrm{d} ^\mathrm{conf}(\mathbf{x})\mathbf{u}_\mathrm{n} + \nu_\mathrm{p} \mathbf{P}\, .
\end{aligned}
\end{equation}
Here, $\mathbf{u}_n$ is the component of $\mathbf{u}$ normal to the circular boundary.
The additional elastic restoring force $- \nu_\mathrm{d} ^\mathrm{conf}(\mathbf{x})\mathbf{u}_\mathrm{n}$ mimics the role of the circular confinement.
Its strength $\nu_\mathrm{d} ^\mathrm{conf}(\mathbf{x})$ depends on the spatial position $\mathbf{x}$. For comparison with the results in Ref.~\cite{xu2023autonomous}, we use the relation $\nu_\mathrm{d} ^\mathrm{conf}(\mathbf{x}) = 2 \nu_{\mathrm{d}0}^\mathrm{conf} %
|\mathbf{x} - \mathbf{x}_0|/d$, where $\mathbf{x}_0$ is the center of the circular system and $d$ marks its diameter. 
The strength of the force thus increases from zero to $\nu_{\mathrm{d}0}^\mathrm{conf}$ when moving from the center to the circular boundary.
When deriving the results shown in the figures of the main article, we set $\nu_{\mathrm{d}0}^\mathrm{conf} = \nu_\mathrm{d} /2$.

Time integration of Eqs.~(1) and (3) in the main article is performed by a fourth-order Runge--Kutta scheme~\cite{suli2003introduction} with time step $\Delta t = 10^{-3}$.
To solve the dynamic equations at every time step, we employ a pseudo-time stepping scheme combined with an artificial compressibility method~\cite{chorin1997numerical}.
Here, we replace Eqs.~(\ref{eq:StokesEquationPlusConfinement}) for an incompressible medium by
\begin{equation}
\label{eq:StokesEquationPseudoTime}
\begin{aligned}
\partial_\tau p &= - \alpha^2 \nabla \cdot \mathbf{v}\, , \\
\partial_\tau \mathbf{v} &= - \nabla p + \eta \nabla^2 \mathbf{v} + \mu \nabla^2 \mathbf{u} - \nu_\mathrm{v}  \mathbf{v} - \nu_\mathrm{d}  \mathbf{u} - \nu_\mathrm{d} ^\mathrm{conf}(\mathbf{x})\mathbf{u}_\mathrm{n} + \nu_\mathrm{p} \mathbf{P}\, ,
\end{aligned}
\end{equation}
where $\tau$ is a pseudo-time variable.
The solution of Eqs.~(\ref{eq:StokesEquationPseudoTime}) approaches the solution of Eqs.~(\ref{eq:StokesEquationPlusConfinement}) for appropriately short pseudo-time stepping $\Delta\tau$~\cite{chorin1997numerical}.
After every Runge--Kutta step, which updates the fields $\mathbf{P}$ and $\mathbf{u}$, we perform $1000$ time steps of $\Delta \tau = 10^{-3}$ using an explicit Euler method~\cite{suli2003introduction}. In this way, we solve Eqs.~(\ref{eq:StokesEquationPseudoTime}) and obtain the updated field $\mathbf{v}$. 
The parameter $\alpha$, often denoted as artificial speed of sound~\cite{chorin1997numerical}, is set to $\alpha = 10$, ensuring that the velocity field is divergence-free after each pseudo-time stepping procedure.

We start our calculations during each evaluation from a quiescent, isotropic state of small random perturbations to the polar order parameter field taken from a uniform distribution over $[-0.01,0.01]$ for the $x$ and $y$ component, respectively.
The investigated circular systems have a diameter of $d=6$, which we resolve via $128\times128$ grid points in our discretized Cartesian coordinates.

\newpage

\bibliography{references}